\documentclass{iaus}
\usepackage{graphicx}

\title[Dust evolution in protoplanetary disks]
{Dust evolution in protoplanetary disks}

\author[J.-F. Gonzalez, L. Fouchet, S.T. Maddison \& G. Laibe]
{Jean-Fran\c{c}ois Gonzalez$^1$, Laure Fouchet$^2$, Sarah T. Maddison$^3$ \and Guillaume Laibe$^1$}

\affiliation{$^1$Universit\'e de Lyon, Lyon, F-69003, France;
Universit\'e Lyon 1, Villeurbanne, F-69622, France;
CNRS, UMR 5574, Centre de Recherche Astrophysique de Lyon,\\
\'Ecole Normale Sup\'erieure de Lyon, 46 all\'ee d'Italie,
F-69364 Lyon cedex 07, France\\
email: {\tt Jean-Francois.Gonzalez@ens-lyon.fr, Guillaume.Laibe@ens-lyon.fr}
\\[\affilskip]
$^2$Department of Physics, ETH Zurich, CH-8093 Zurich, Switzerland\\
email: {\tt fouchet@phys.ethz.ch}
\\[\affilskip]
$^3$Centre for Astrophysics and Supercomputing, Swinburne University
of Technology,\\ PO Box 218, Hawthorn, VIC 3122, Australia\\
email: {\tt smaddison@swin.edu.au}}

\pubyear{2008}
\volume{249}
\jname{Exoplanets: Detection, Formation and Dynamics}
\editors{Y.-S. Sun, S. Ferraz-Mello \& J.-L. Zhou, eds.}
\begin{document}

\maketitle

\begin{abstract}
We investigate the behaviour of dust in protoplanetary disks under the action of gas drag using our 3D, two-fluid (gas+dust) SPH code. We present the evolution of the dust spatial distribution in global simulations of planetless disks as well as of disks containing an already formed planet. The resulting dust structures vary strongly with particle size and planetary gaps are much sharper than in the gas phase, making them easier to detect with ALMA than anticipated. We also find that there is a range of masses where a planet can open a gap in the dust layer whereas it doesn't in the gas disk. Our dust distributions are fed to the radiative transfer code MCFOST to compute synthetic images, in order to derive constraints on the settling and growth of dust grains in observed disks.
\end{abstract}

\firstsection % if your document starts with a section,
\section{Introduction}
\label{SectIntro}

Dust grains in disks around young stars are thought to be the building
blocks of planets \cite[(Dominik \etal\ 2007)]{Dominik2007}. They grow via
sticking in low-velocity collisions from (sub)micron sizes up to decimetric
sizes. The higher collision velocities of larger grains makes them shatter
upon impact and prevents them from growing further. Yet we know that they
must reach planetesimal sizes. How they overcome the decimetric barrier is
the subject of much debate. Part of the solution to this problem could be
found in the reduction of collision velocities via the increase of the dust
layer density. This can be achieved by the settling of dust to the disk
midplane.

Dust is also essential to the interpretation of disk observations. Indeed, the
dust opacity largely dominates that of the gas, except in particular lines of
molecular gas, which are more difficult to detect than dust thermal emission.
Therefore, most images of disks trace the spatial distribution of dust, not
that of the gas. Both components are often assumed to be well mixed,
with a uniform gas-to-dust mass ratio of 100, but, as discussed in
Sect.~\ref{SectSettling}, this is not always the case and one has to be
careful when deriving gas distributions from observations of dust in
circumstellar disks.

We present here our work on dust evolution in protoplanetary disks using 3D
SPH simulations. We treat vertical settling and radial migration of dust in
planetless disks in Sect.~\ref{SectSettling} and present the resulting
synthetic images and comparison to observations in Sect.~\ref{SectImages}.
We describe the variety of structures of disks with gap-forming planets in
Sect.~\ref{SectGaps} and our first treatment of grain growth in
Sect.~\ref{SectGrowth}.

\section{Settling and migration}
\label{SectSettling}

In a non self-gravitating disk, gas feels the gravity of the central star and
its own pressure gradient. As a result, it will orbit at sub-keplerian
velocities. On the other hand, dust grains do not interact with each other
and only feel the star's gravity: they have keplerian orbits. In a disk with
both components, this velocity difference gives rise to aerodynamic drag,
which slows down the dust and makes it migrate radially towards the central
star. In the vertical direction, drag makes the dust settle towards the
midplane.

Analytical studies of the radial migration of solid particles have been done
by \cite[Weidenschilling (1977)]{Weidenschilling1977} and of their vertical
settling by \cite[Garaud \etal\ (2004)]{Garaud2004}. In order to simultaneously
study both effects, we developped a 3D, bi-fluid (gas+dust), SPH code to model
vertically isothermal, non self-gravitating dusty disks
\cite[(Barri\`ere-Fouchet \etal\ 2005)]{BF05}. The two inter-penetrating
phases representing gas and dust interact via aerodynamic drag. Grain growth
is not taken into account at this point (see Sect.~\ref{SectGrowth}).

\begin{figure}[t]
\begin{center}
\includegraphics[width=4.5cm]{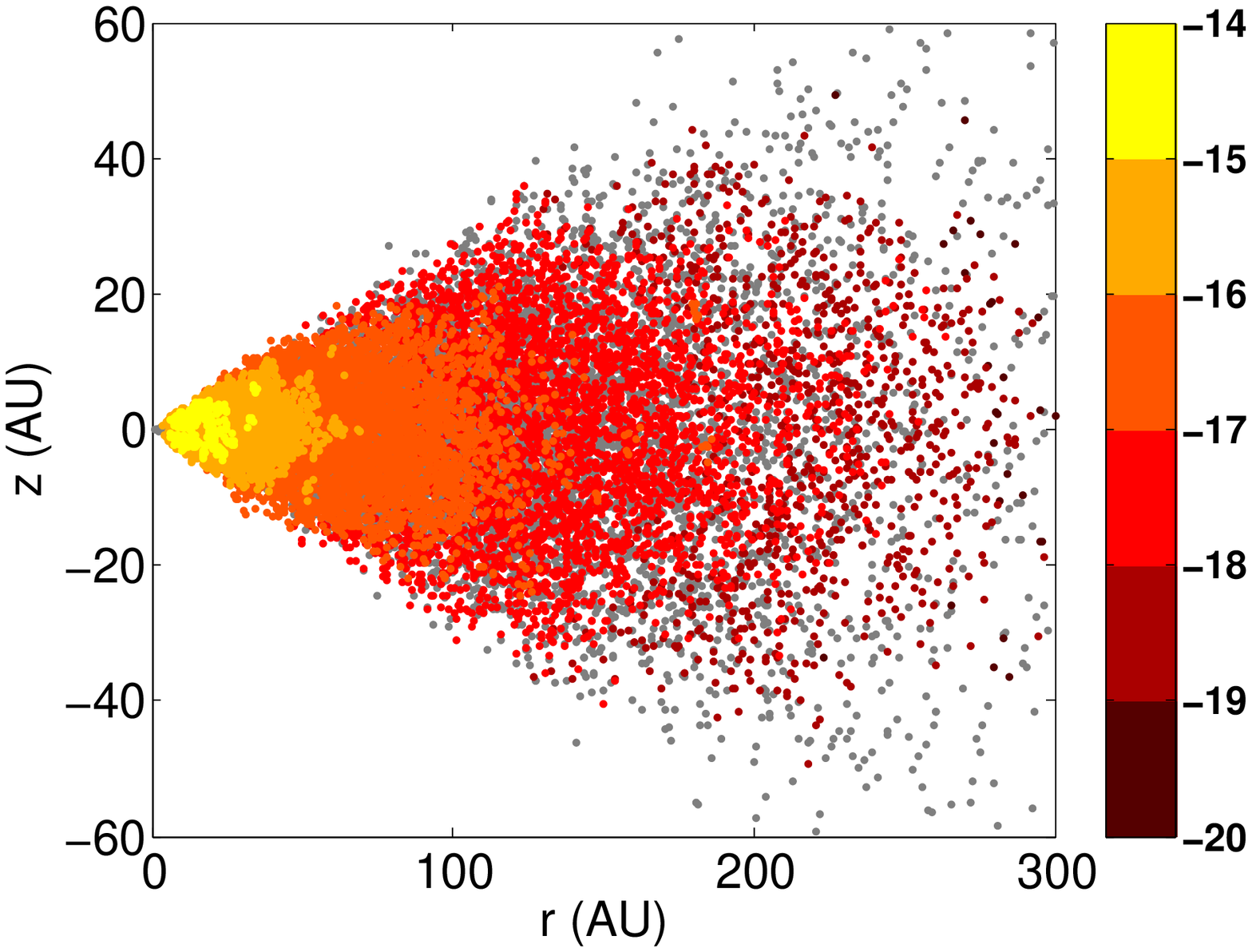}%
\includegraphics[width=4.5cm]{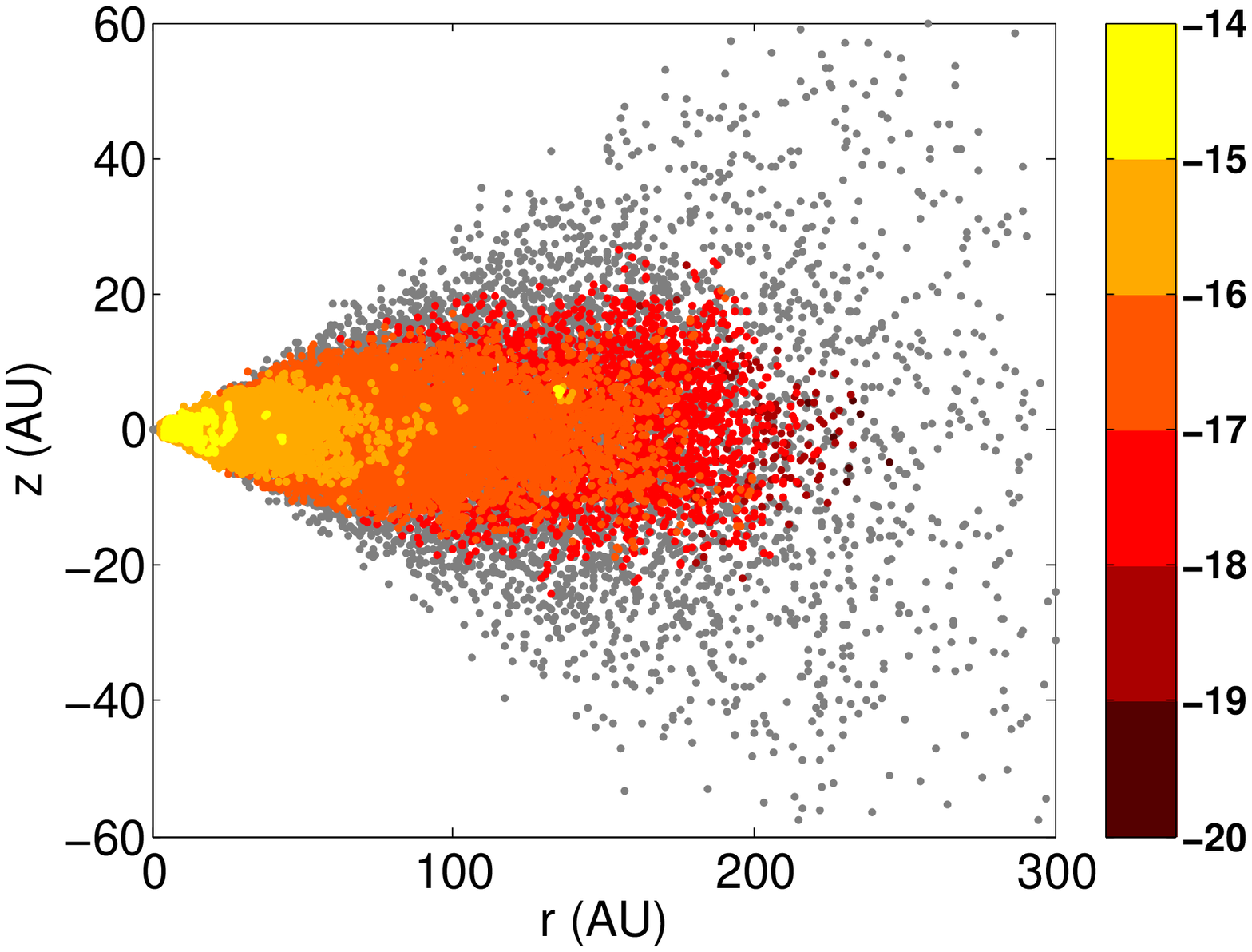}%
\includegraphics[width=4.5cm]{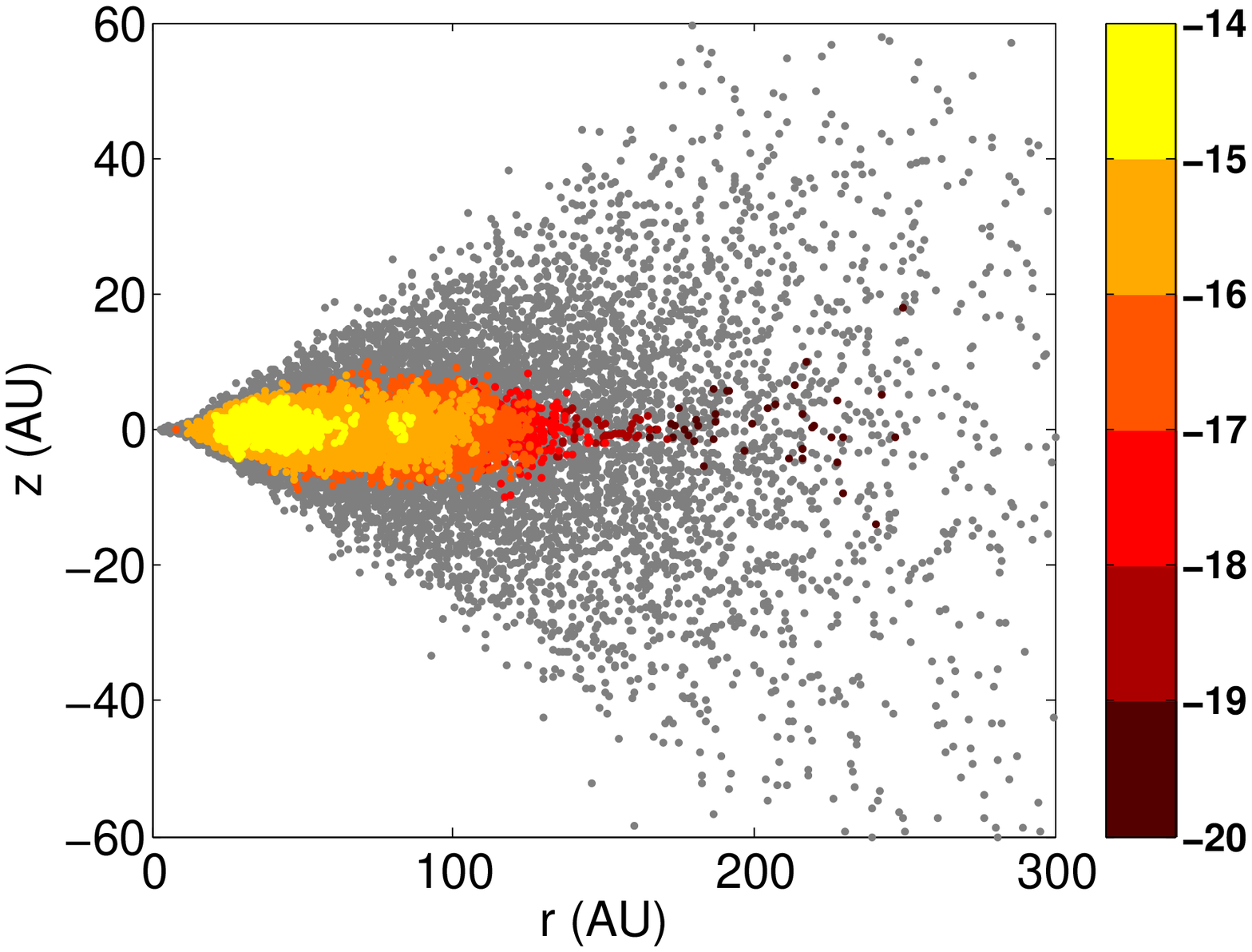}
\includegraphics[width=4.5cm]{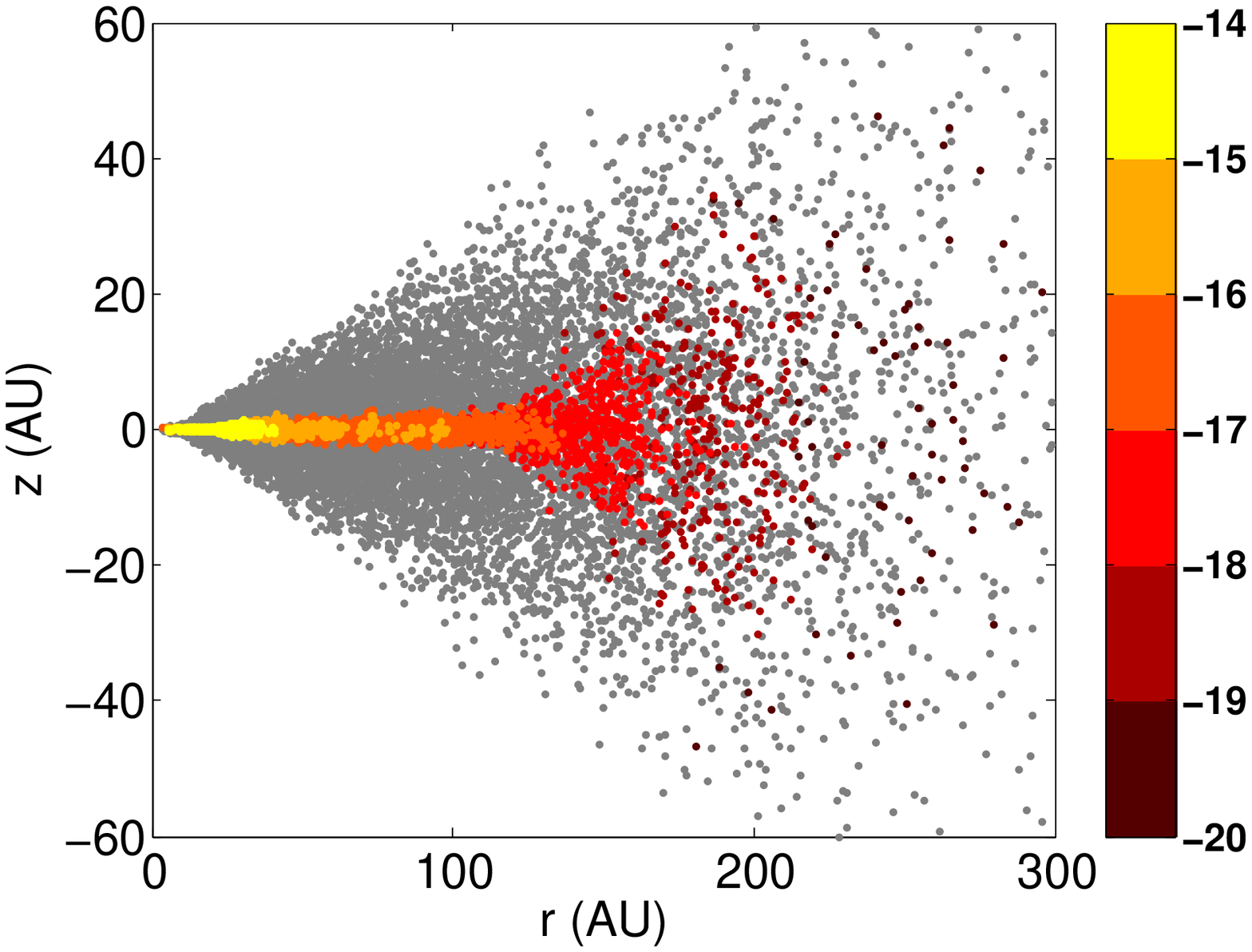}%
\includegraphics[width=4.5cm]{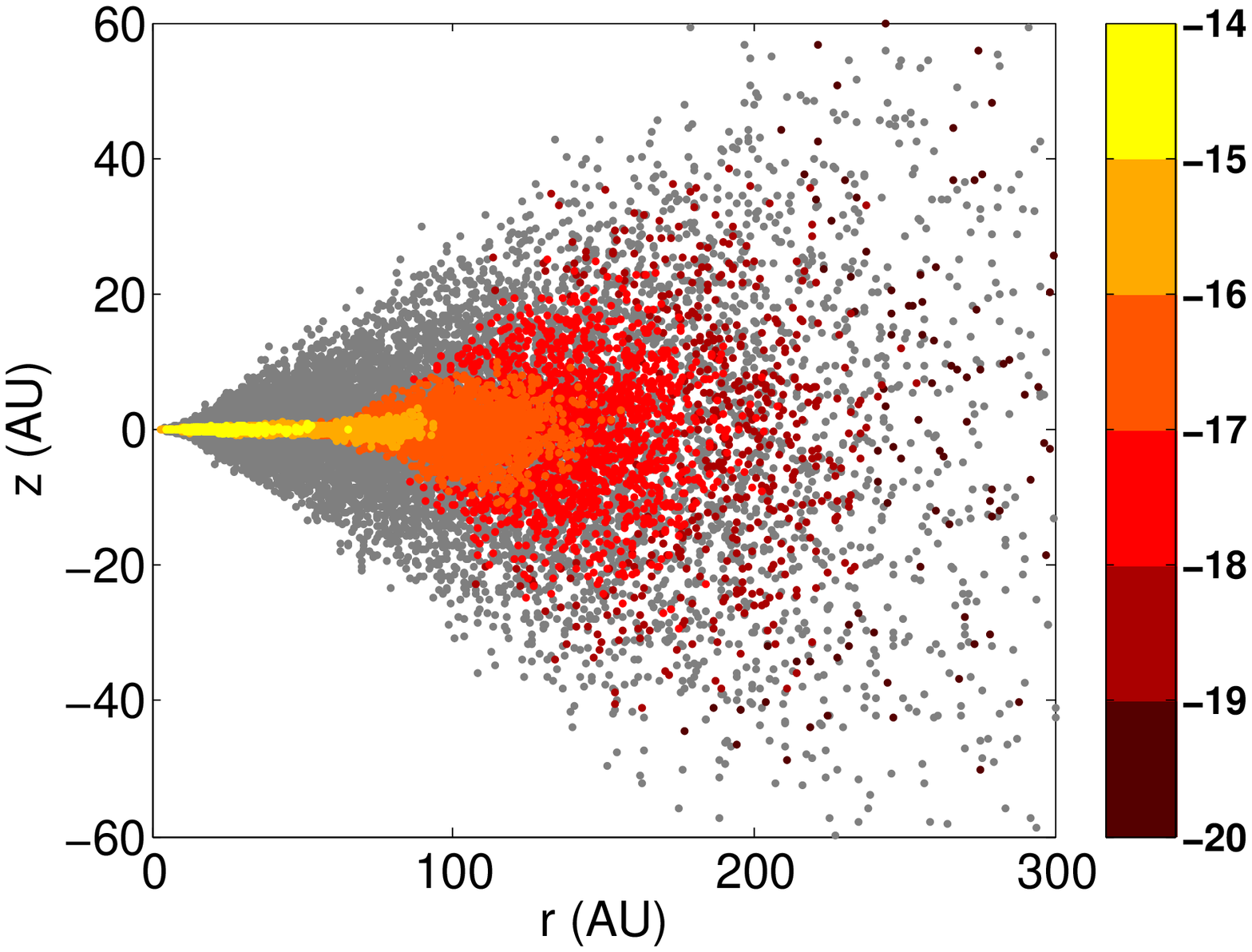}%
\includegraphics[width=4.5cm]{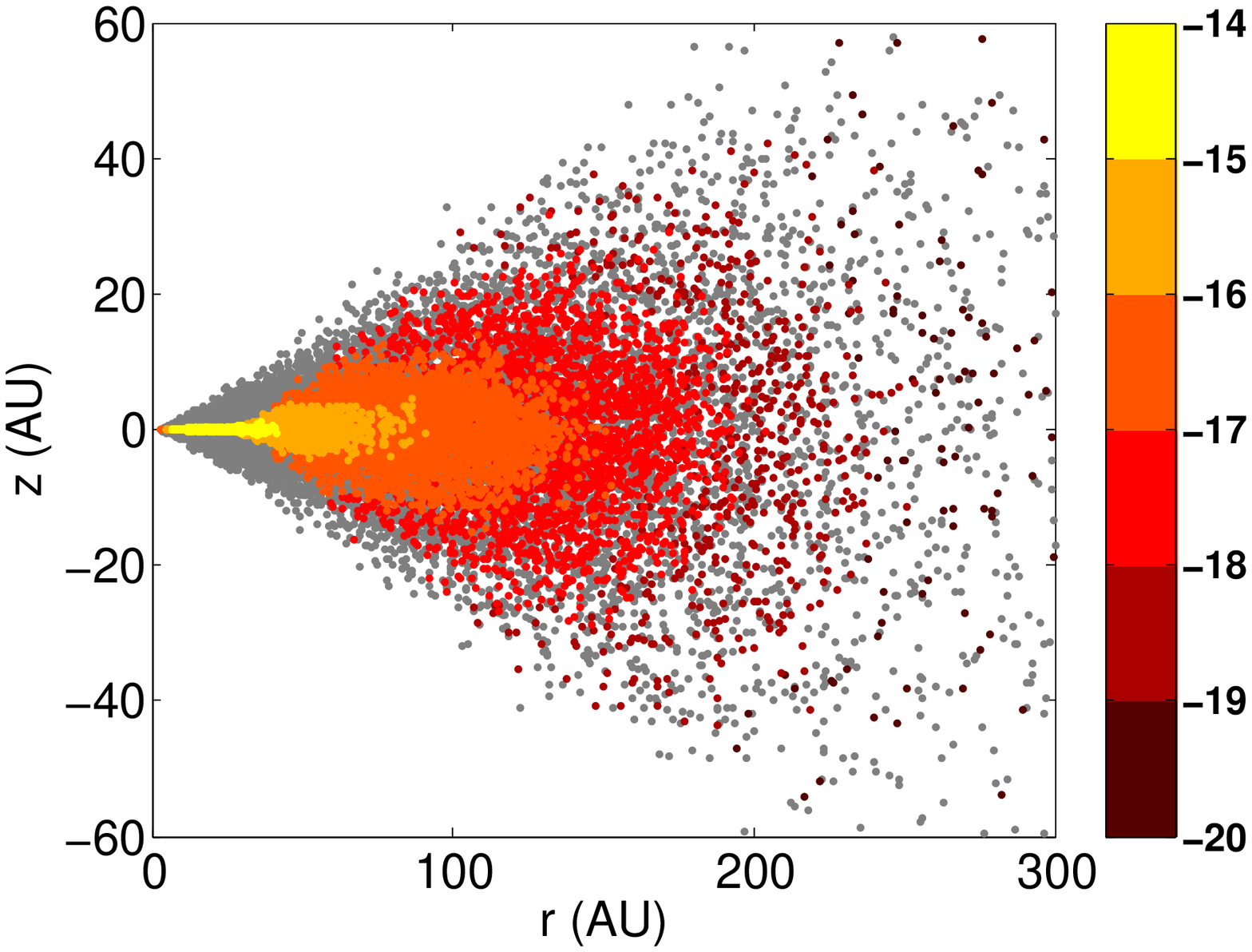}
\caption{Gas (grey) and dust (colour represents $\log\rho_\mathrm{dust}$,
in g\,cm$^{-3}$) distributions in our CTTS disk viewed edge-on. Dust grain
sizes are, from left to right and top to bottom: 10~$\mu$m, 100~$\mu$m, 1~mm,
1~cm, 10~cm, 1~m.}
\label{FigSettling}
\end{center}
\end{figure}

We ran simulations for a $0.01\ M_\odot$ disk in orbit around a 1~$M_\odot$
star, representative of a classical T~Tauri star (CTTS) disk, and composed of
99\% gas and 1\% dust in mass. We obtained the resulting dust spatial
distributions for a set of grain sizes ranging from 1~$\mu$m to 10~m
\cite[(Barri\`ere-Fouchet \etal\ 2005)]{BF05}, a
selection of which is illustrated in Fig.~\ref{FigSettling}.

We found that for small sizes (1--10~$\mu$m), with the strongest drag force,
the dust is so strongly coupled to the gas that it follows its motion and both
components are well mixed. On the other hand, the drag force for large grain
sizes (1--10~m) is weak and the dust is almost insensitive to the gas, it
occupies the whole disk again. It is for the median sizes (100~$\mu$m--10~cm)
that all the interesting dynamics happens and the gas drag has a strong
influence on the resulting distributions, showing important vertical settling
in the inner regions and depletion in the outer disk due to inwards radial
migration, with varying strength depending on grain size. 

We would like to point out that the efficiency of both processes depends not
only on grain size but also on the nebula parameters, via the densities.
Therefore, the size of the fastest inwards migrating grains is not universal.
It is commonly thought to always be of the order of 1~m, as found by
\cite[Weidenschilling (1977)]{Weidenschilling1977} for the minimum mass solar
nebula (MMSN), whereas for our CTTS disk, we find it to be smaller, around
1~mm to 1~cm, and to depend on the position in the disk.

\section{Synthetic images}
\label{SectImages}

Numerous scattered light images of protoplanetary disks are now available
\cite[(Watson \etal\ 2007)]{Watson2007}.  Light scattering depends on grain
size and, because of the wavelength dependency of opacity, observations at
different wavelengths reveal different depths in the disk. As a result,
multi-wavelength observations in scattered light constitute a probe of dust
settling.

In order to link our hydrodynamical simulations to observations of disks, we
used our resulting dust distributions as an input to MCFOST, a continuum 3D
Monte-Carlo radiative transfer code \cite[(Pinte \etal\ 2006)]{Pinte2006}.
It can produce synthetic scattered light images, as well as spectral energy
distributions (SEDs) and polarization maps. We found that dust settling and
migration cause wavelength variations of the disk apparent size and of the
dark lane width, in agreement with observations. They also affect SEDs,
causing a change of slope between 10 and 200~$\mu$m. However, these effects
are small, and a single observation would not give strong indications on the
amount of settling. It is with a combination of multi-wavelength, multi-technique observations that one can hope to efficiently constrain the
dust distribution.

\begin{figure}[t]
\begin{center}
\includegraphics[width=12cm]{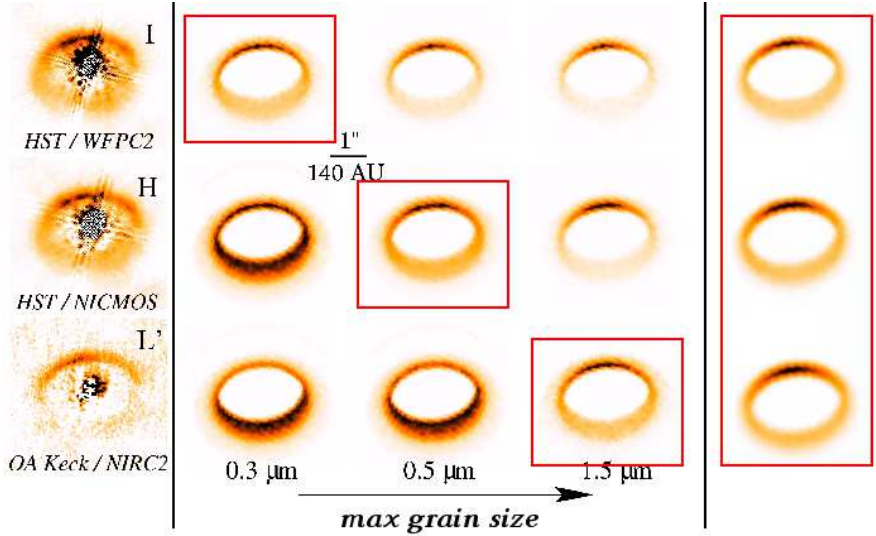}
\caption{Observations (left panel) and synthetic images for models without
(center panel) and with dust settling (right panel) of GG~Tau's circumbinary
ring in the I, H and L' bands.}
\label{FigGGTau}
\end{center}
\end{figure}

This is beautifully illustrated in the case of GG~Tau. The left panel of
Fig.~\ref{FigGGTau} shows three scattered light images of its circumbinary
ring observed in the I, H, and L' bands. The middle panel displays the
corresponding synthetic images for three models without dust settling,
therefore with gas and dust well mixed in the disk, for which the dust size
distribution (with the same slope as that of the interstellar medium), is
truncated at a maximum size of 0.3, 0.5 and 1.5~$\mu$m, from left to right.
No single model is able to simultaneously reproduce the observations in all
three bands, and in particular the contrast between the front and back sides
of the ring, but each one seems appropriate for a single band. This points
towards a variation of the dust size distribution with depth in the disk.
The right panel of Fig.~\ref{FigGGTau} shows synthetic images produced from
new hydrodynamical simulations of GG~Tau's circumbinary ring including dust
settling and migration. Without any fit to the data, the agreement with
observations is now much better, showing that dust had indeed settled in this
ring. Polarization data comforting this result, as well as more detailed
azimuthal brightness profile comparisons of observations and models, are
presented in \cite[(Pinte \etal\ 2007)]{Pinte2007}. The remaining
discrepancies are likely due to missing physics in our simulations, among
which grain growth.

\section{Planetary gaps}
\label{SectGaps}

In a disk containing a planet, gravitational perturbations of the planet cause
density waves, leading to an exchange of angular momentum with the disk. The
planet pushes the disk exterior (resp. interior) to its orbit further out
(resp. in), therefore opening a gap. The gap is sustained if there is an
equilibrium between tidal torques, acting to clear the gap, and viscous
torques, acting to fill it. The observation of such a gap in a disk would
provide a clear indication of the presence of a planet. Simulations of
gaps in 2D gas disks by \cite[Wolf \& D'Angelo (2005)]{Wolf2005} showed
that a 1~$M_\mathrm{J}$ planet can open a gap that will be detectable by ALMA
at sub-mm wavelengths up to a distance of 100~pc. In 2D simulations of dusty
disks, \cite[Paardekooper \& Mellema (2006)]{PM2006} found that a smaller
planet mass is needed to open a gap in the dust phase than in the gas.

\begin{figure}[t]
\begin{center}
\includegraphics[width=13cm]{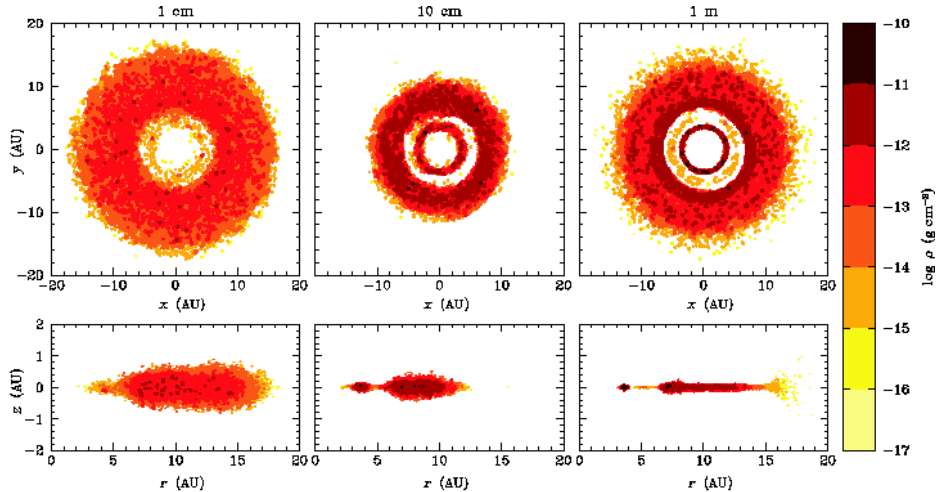}
\caption{Gap created in the dust phase of the MMSN disk for 1-cm, 10-cm and
1-m grains, from left to right, by a 1~$M_\mathrm{J}$ planet at 5.2~AU.
Top panel: face-on view, bottom panel: edge-on view.}
\label{FigGapSize}
\end{center}
\end{figure}

We ran new 3D simulations of dusty disks with an embedded planet for two
configurations, the CTTS disk we modeled previously and a MMSN disk, in which
we vary the grain size and planet mass. We found that gap formation is much
more rapid and striking in the dust layer rather than in the gas disk, and even
more so than in 2D simulations \cite[(Maddison \etal\ 2007; Fouchet \etal\
2007)]{Maddison2007,Fouchet2007}. Additionally, the structures caused in the
dust phase depend strongly on the grain size, as illustrated in
Fig.~\ref{FigGapSize}, due to varying drag and radial drift velocity
throughout the disk.

\begin{figure}[t]
\begin{center}
\includegraphics[width=6.5cm]{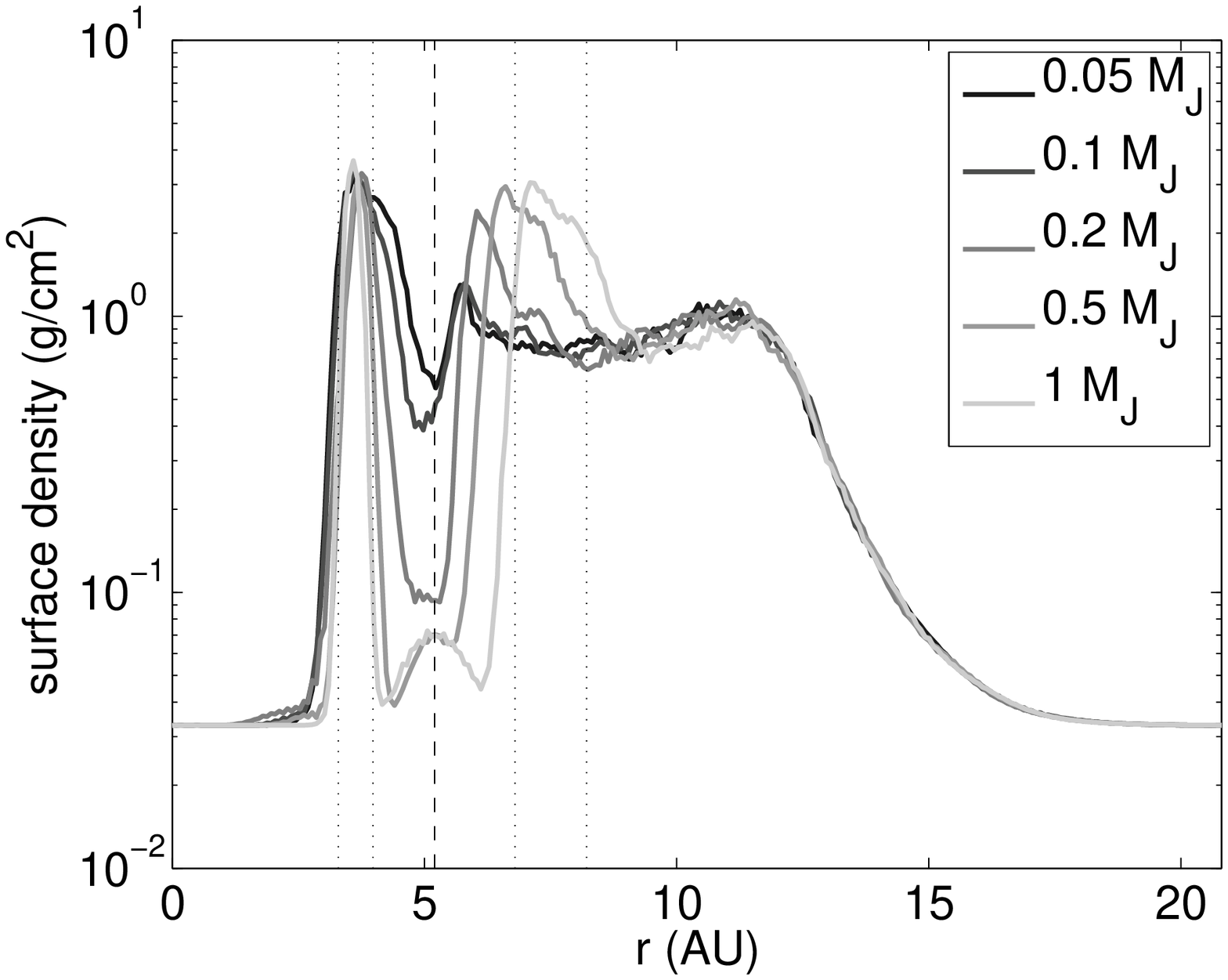}%
\includegraphics[width=6.5cm]{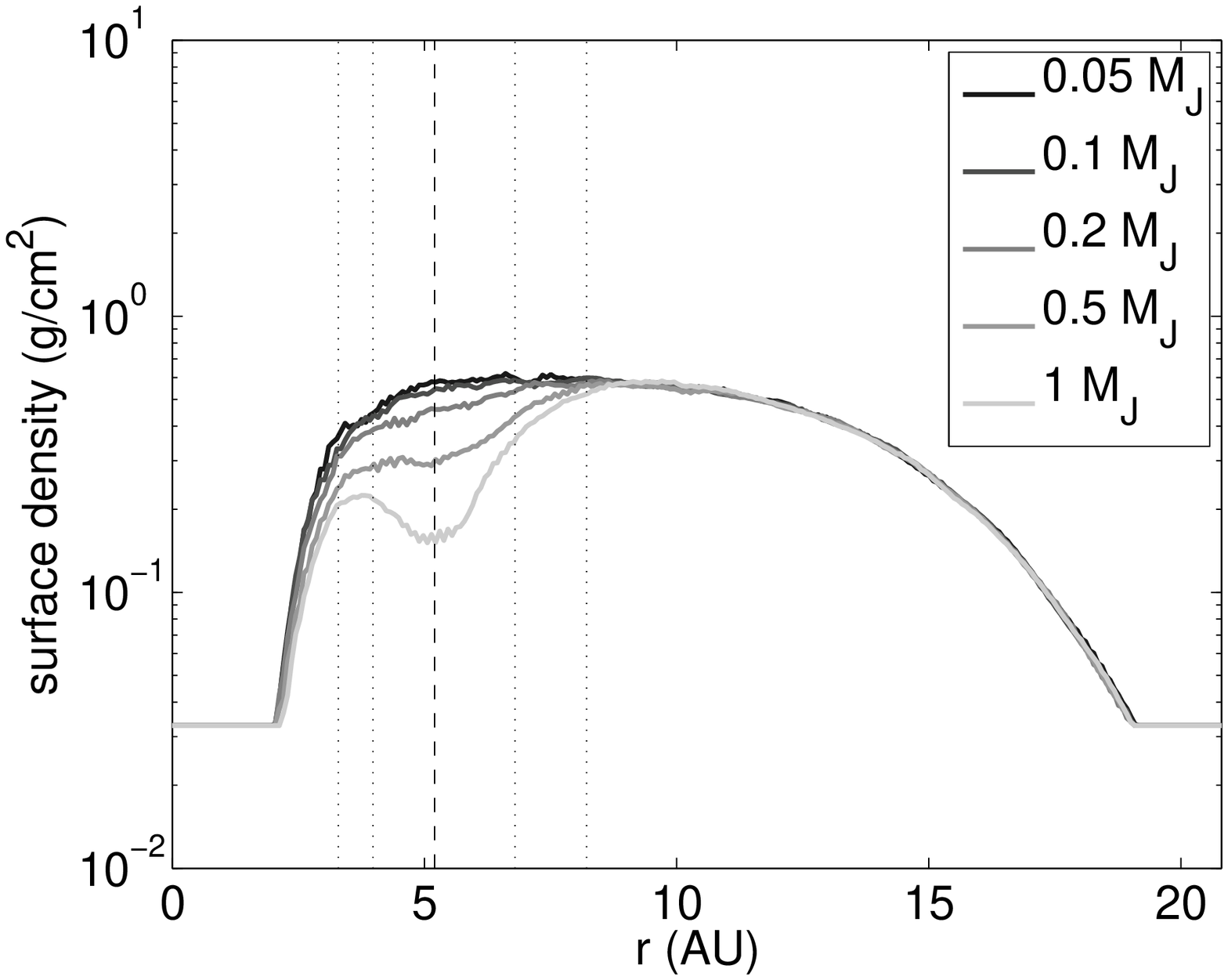}
\end{center}
\caption{Azimuthally averaged surface density profile of the MMSN disk for the
dust (1-m boulders, left) and the gas (right) with a planet of varying mass
at 5.2~AU. The gas density is divided by 100 for direct comparison to the dust
density.}
\label{FigGapSigma}
\end{figure}

Gaps are more prominent as the planet mass increases. Since they are also much
deeper in the dust than in the gas, much lighter planets are needed to open
one in the dust phase, the one visible in most disk observations, than in the
gas phase. In our MMSN models, we found that a 0.2~$M_\mathrm{J}$ planet can
open a gap in the dust layer while only a small surface density dip is
observed in the gas (see Fig.~\ref{FigGapSigma}). Lighter planets down to
0.05~$M_\mathrm{J}$ create density dips in the dust, with steep gradients on
both sides that should be visible by ALMA \cite[(Varni\`ere \etal\
2006)]{Varniere2006}. This suggests that planet gaps will be much easier to
detect, with a lower threshold on planet mass, than previously anticipated.

Since the influences of grain size, affecting the disk thickness and radial
profile (Fig.~\ref{FigGapSize}) and of planet mass, affecting the gap
structure (Fig.~\ref{FigGapSigma}) are different (see also \cite[Fouchet
\etal\ 2007]{Fouchet2007} for more detail), future observations of planet gaps
will therefore allow to constrain the dust grain size distribution in addition
to the planet mass.

\begin{figure}[t]
\begin{center}
\includegraphics[width=12cm]{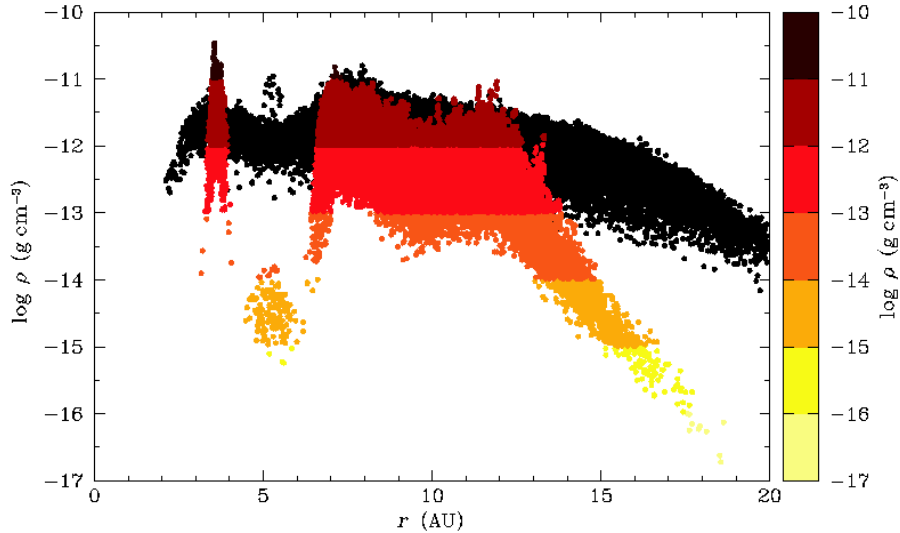}
\end{center}
\caption{Volume density profile of the gas (black) and dust (colour) in the
MMSN disk with a 1~$M_\mathrm{J}$ planet at 5.2~AU for 1-m boulders.}
\label{FigGapRho}
\end{figure}

Gaps can even play a role in planet formation. Because of a very efficient
vertical settling, the volume density of the dust reaches that of the gas in
several regions of the disk, and even exceeds it at the gap inner edge
(see Fig.~\ref{FigGapRho}). This affects the dynamics in that region and can
favour planetesimal growth, ultimately leading to the formation of a second
planet in the vicinity of the first one.

\section{Grain growth}
\label{SectGrowth}

We recently started to add grain growth in our code by implementing a scheme
able to treat the variation of grain sizes via an analytical prescription.
We tested it with the simple model of
\cite[Stepinski \& Valageas (1997)]{Stepinski1997}, who grow solid particles
made of water ice by sticking, without taking shattering into account. We
found that grain growth is very efficient, especially in the denser inner
regions of the disk, but is too fast, as can be expected
\cite[(Dullemond \etal\ 2005)]{Dullemond2005}. For details, see the poster
by Laibe, this volume, and references therein. More extensive tests are
presented in \cite[Laibe \etal\ (2008)]{Laibe2008}.

\section{Conclusion}
\label{SectConcl}

Our results demonstrate that, with or without an embedded planet, gas and
dust are not well mixed in protoplanetary disks. The density enhancements
we find in the dust component shape their observed images, and help solid
particles to aggregate.

Similarly to our work on GG~Tau, we have started to compute synthetic images
of disks with gaps in order to make better quantitative predictions of what
ALMA will be able to detect. Gas dispersal, not taken into account here, can
lower the drag force and alter the dust distribution, and should be considered
when interpreting observations.

Now that we have implemented in our code a mechanism to treat grain growth,
we are working on an improved and more realistic physical model, taking into
account detailed microscopic processes, as well as shattering, which produces
small grains via collisional cascade. Our goal is to be able overcome the
decimetric barrier. Our approach is different, but complementary to the one
of \cite[Johansen \etal\ (2007)]{Johansen2007}, who find a concentration of
solid particles in transient high pressure regions.

A physically consistent model of grain growth needs a proper implementation
of turbulence in the gas, because it determines the kinematics of grains.
As it is unfeasible to include a full treatment of turbulence in our code,
we started working on a simplified, but realistic description.

\end{document}